\documentclass[11pt,a4paper]{article}
\usepackage[utf8]{inputenc}
\usepackage[T1]{fontenc}
\usepackage{lmodern}
\usepackage{microtype}
\usepackage[margin=1.95cm]{geometry}
\usepackage{setspace}
\usepackage{parskip}
\usepackage{graphicx}
\usepackage{wrapfig}
\usepackage{booktabs}
\usepackage{enumitem}
\usepackage{hyperref}
\hypersetup{colorlinks=true,linkcolor=blue,citecolor=blue,urlcolor=blue}
\usepackage{titling}
\usepackage{fancyhdr}
\usepackage{caption}
\usepackage{amsmath,amssymb}
\usepackage{titlesec}

\usepackage[
  backend=biber,
  style=numeric,
  sorting=none,
  giveninits=true,
  maxnames=1
]{biblatex}
\defbibheading{bibliography}[\refname]{\section*{#1}\vspace{-8pt}}

\titleformat{\section}
  {\large\normalfont\bfseries}
  {\thesection}{1em}{}
  
\renewbibmacro*{in:}{}

\AtEveryBibitem{%
  \clearfield{title}%
  \clearfield{subtitle}%
  \clearfield{number}%
  \clearfield{issue}%
  \clearfield{month}%
  \clearfield{day}%
  \clearfield{doi}%
  \clearfield{url}%
  \clearfield{isbn}%
  \clearfield{issn}%
  \clearfield{note}%
  \clearfield{addendum}%
  \clearfield{howpublished}%
  \clearfield{eprintclass}% kills [gr-qc]
}

\renewbibmacro*{date}{\printfield{year}}

\DeclareFieldFormat{eprint}{arXiv:#1}

\DeclareBibliographyDriver{article}{%
  \printnames{author}\setunit{\addspace}%
  \printfield{year}\setunit{\addcomma\space}%
  \printfield{journaltitle}\setunit{\addcomma\space}%
  \printfield{volume}\setunit{\addcomma\space}%
\iffieldundef{journaltitle}
  {\iffieldundef{eprint}{}{%
     \setunit{\addcomma\space}\printfield{eprint}}}
  {}%
\finentry
}

\DeclareBibliographyDriver{online}{%
  \printnames{author}\setunit{\addspace}%
  \printfield{year}%
  \iffieldundef{eprint}{}{%
    \setunit{\addcomma\space}\printfield{eprint}}%
  \finentry
}
\DeclareBibliographyDriver{unpublished}{\usebibmacro{bibindex}\usebibmacro{begentry}\usebibmacro{author/translator+others}\setunit{\addspace}\printfield{year}\iffieldundef{eprint}{}{\setunit{\addcomma\space}\printfield{eprint}}\usebibmacro{finentry}}

\defbibenvironment{bibliography}
  {}
  {}
  {\addspace
   \printtext[labelnumberwidth]{%
     \printfield{prefixnumber}%
     \printfield{labelnumber}}%
   \addhighpenspace}

\bibliography{bib}

\pretitle{\vspace*{-1cm}\begin{center}\LARGE\bfseries}
\posttitle{\par\end{center}\vskip 0.5em}
\preauthor{\begin{center}\large}
\postauthor{\end{center}}
\predate{\begin{center}\large}
\postdate{\end{center}}

\begin{document}

% -----------------------
% Cover page
% -----------------------
\begin{titlepage}
  \centering
  \vspace*{1cm}
  {\Huge\bfseries Expanding Horizons \\[6pt] \Large Transforming Astronomy in the 2040s \par}
  \vspace{1.5cm}

  % Title (editable)
  {\LARGE Luminous Fast Blue Optical Transients in the 2040s}\par
  \vspace{1cm}

  % Metadata block
  \begin{tabular}{p{4.5cm}p{11cm}}
    \textbf{Scientific Categories:} & Time-domain, Black Holes, Supernovae, Instrumentation \\
    %(e.g. Cosmology; Time-domain; Instrumentation) \\
    \\
    \textbf{Submitting Author:} & Name: Ashley Chrimes \\
    & Affiliation:  Radboud University (NL) \& ESA-ESTEC \\
    & Email: a.chrimes@astro.ru.nl \\
    \\
    \textbf{Contributing authors:} & \\
    %List of co-authors (name, affiliation, email). Up to first three authors should be in an ESO Member State. \\
     Nikhil Sarin & University of Cambridge (UK), ns2043@cam.ac.uk\\
     Deanne Coppejans & University of Warwick (UK), deanne.coppejans@warwick.ac.uk\\
     Paul Groot & Radboud University (NL), p.groot@astro.ru.nl\\
     Anne Inkenhaag & University of Bath (UK), ai707@bath.ac.uk\\
     Peter Jonker & Radboud University (NL), p.jonker@astro.ru.nl\\
     Tom Killestein & University of Warwick (UK), thomas.killestein@warwick.ac.uk\\
     Dan Perley & Liverpool John Moores University (UK), D.A.Perley@ljmu.ac.uk\\
     Miika Pursiainen & University of Warwick (UK), miika.pursiainen@warwick.ac.uk\\
  \end{tabular}

  \vspace{1cm}

\end{titlepage}

% -----------------------
% Main sections (match DOCX template)
% -----------------------

%https://next.eso.org/
  \textbf{Abstract:}
  %\vspace{0.5em}
  Luminous Fast Blue Optical Transients (LFBOTs) are a class of extragalactic transient of uncertain origin. Several hypotheses have been put forward which could feasibly be consistent with the sample number of events discovered thus far, including tidal disruption events around intermediate mass black holes, failed supernovae and mergers of stars with black holes. Whatever their origin, it is clear that better understanding LFBOTs will provide unique insight into the black hole formation/growth, central engine physics and transient-host galaxy interactions - themes which are expected to drive research in transient astronomy over the coming decades. The vast majority of LFBOTs are missed by current photometric surveys, or not efficiently selected for detailed follow-up. This white paper outlines the observing capabilities required on a 2040 timescale to maximise the discovery potential from these enigmatic events.

\section{Introduction and Background}
\label{sec:intro}
%Short introduction to the science question. Provide context, current state of the field, and motivation for the 2040s vision.  
%\emph{(Keep the science part to less than 3 pages where possible.)}

%https://next.eso.org/call-for-white-papers/ - upload link

Wide-field, high-cadence sky surveys have continuously been expanding the parameter space of known transients over the past decade. Fast, blue optical transients (FBOTs) were first identified as a class in the Pan-STARRS1 Medium Deep Survey by \cite{2014ApJ...794...23D}. Loosely, `fast' here refers to timescales of $>$0.1-0.3\,mag\,day$^{-1}$. 

Despite a decade of efforts to improve the efficiency of their detection and follow-up, it remains challenging to detect these events in large numbers with current facilities, and even more challenging to follow-up with spectroscopy, which is crucial to understand their natures. In particular, the origin of the Luminous FBOTs (LFBOTs, the most luminous and fastest member of the class) remains unclear. 
%%%%%%%%%%%%%%%%%%%%%%%%%%%%%%%%%%%%

% LFBTOs>
The first LFBOT to be identified was AT2018cow \cite{2018ApJ...865L...3P}, discovered by ATLAS, which surveys the observable sky with a cadence of 2 days and down to magnitude $m\sim19$ \cite{2018PASP..130f4505T}. It was reported as an unusual transient due to the large $\Delta$m between the discovery epoch and the previous upper limit at its location, indicating a rapid rise ($>$1.5\,mag/day) and high luminosity (given the low redshift, $z=0.014$, of the host galaxy). Subsequent photometry (from 1.7 days post discovery) and spectroscopy (from 2.6 days) revealed a transient with a featureless, hot ($\sim$30000K) blackbody spectrum, a peak absolute magnitude similar to the most luminous supernova, but a rapid rise and fading timescale which placed strong constraints on the ejecta mass \cite{2018ApJ...865L...3P}. Broad H and He lines emerged after $\sim$2 weeks \cite{2019MNRAS.484.1031P}.

The transient was accompanied by luminous X-ray, submm and radio emission. The short-timescale variability (including possible quasi-periodic oscillations \cite{2022NatAs...6..249P}) and steep decay of the X-ray emission implicated a compact object (neutron star or black hole) central engine \cite{2019ApJ...872...18M}. Continued monitoring found that the transient lightcurve plateaued after one year \cite{2022MNRAS.512L..66S,2023ApJ...955...43C,2023MNRAS.525.4042I,2024ApJ...963L..24M}, and remains hot and slowly fading to this day \cite{2025MNRAS.544L.108I}. This late-time slow-down in the fading (observed in the UV and X-ray) has been interpreted as evidence for a massive accretion disc around a black hole of several tens to several tens of thousands of Solar masses.

Suggested models for LFBOTs (sometimes referred to as Cow-like transients) include magnetar-powered, ultra-stripped supernovae, failed supernova, tidal disruption events (of white dwarfs around intermediate mass black holes) and black hole-stellar mergers \cite{2019MNRAS.487.2505K,2019ApJ...872...18M,2019MNRAS.484.1031P,2022ApJ...932...84M}. Since ÀT2018cow, LFBOTs have been discovered at a rate of 1--2 per year, with an estimated volumetric rate of 0.1\% of the core-collapse supernova rate \cite{2023ApJ...949..120H}. Other key observations include,
\begin{itemize}
    \item Non-relativistic to mildly relativistic non-jetted outflows \cite{2020ApJ...895L..23C,2023MNRAS.521.3323M,2025MNRAS.537.3298P}. Highly blueshifted (v=0.1c) lines were observed in the LFBOT CSS161010 \cite{2024ApJ...977..162G}.
    \item Dense circumstellar media close to the progenitor, denser even than expected for strong winds from Wolf-Rayet stars \cite{2019ApJ...872...18M,2025ApJ...993L...6N}.
    \item Minute-long optical flares with SN luminosities observed following AT2022tsd \cite{2023Natur.623..927H}. No flares were found in AT2024wpp despite an extensive search \cite{2025ApJ...993...76O}.
    \item A preference for star-forming, low metallicity environments; AT2023fhn was found at a relatively large host galaxy offset \cite{2024MNRAS.527L..47C}.
    \item A near-infrared (NIR) excess at 1--2 weeks, seen in AT2018cow/AT2024wpp \cite{2023ApJ...944...74M,2025MNRAS.537.3298P}. No NIR spectroscopy has been obtained at this phase. 
\end{itemize}

A few key features define LFBOTs as a class - the short rise/fall time, blue colour, X-ray and radio emission. Beyond this, it is not clear that all LFBOTs share the same progenitor. Systematic discovery, a large number of events and efficient, uniform, multi-wavelength follow-up are needed to make progress in this regard. The low ejecta mass and energetic outflows of LFBOTs allow us to uniquely observe nascent compact objects and their associated accretion and outflows, which means they are important targets for answering key questions even beyond the field of transients.
%Other sources can also appear as fast, blue transients, and their efficient detection and follow-up also holds great promise for several areas of astrophysics. These include young supernovae, supernovae with nebular emission (circumstellar interaction) and kilonovae. 

%%%%%%%%%%%%%%%%%%%%%%%%%%%%%%%%%%%%%%%%%%%%%%%%%%%%%%%%%%%%%%%%%%%%%%%%%%%%%%%%%%%%%%%%%%%%%%%%%%%%%%%%%%%%%%%%%%%%%%%%%%%%%%%%%%%%%%%%%%%%%%%%%%%

\section{Open Science Questions in the 2040s}
\label{sec:openquestions}

% State of the field in 2040s
Key questions expected to drive the field in the 2040s, with a connection to LFBOTs, include:
\begin{itemize}
    \item How does the formation, growth and merger of black holes across all mass scales proceed across cosmic history? 
    \item How does the variety of transient phenomena observed map onto the myriad outcomes of stellar evolution? 
    \item What are the central engines of transients and how does accretion couple to / produce the outflows and jets which drive energetic transients?
    \item What role do different transients play in shaping their host galaxies, both mechanically and chemically?
\end{itemize}
For example, there is growing consensus that LFBOTs are powered by BH accretion, but the mass scale of the BHs in question is still uncertain. If LFBOTs are TDEs, they provide a unique method of probing the elusive intermediate mass BHs. If stellar mass BHs are involved, LFBOTs may instead probe `failed' GW sources (i.e. star-BH mergers \cite{2025arXiv251009745K}) and/or the physics of high-mass core-collapse events (e.g. \cite{2025arXiv251003402C}).

\begin{figure*}
        \centering
	\includegraphics[width=0.47\textwidth]{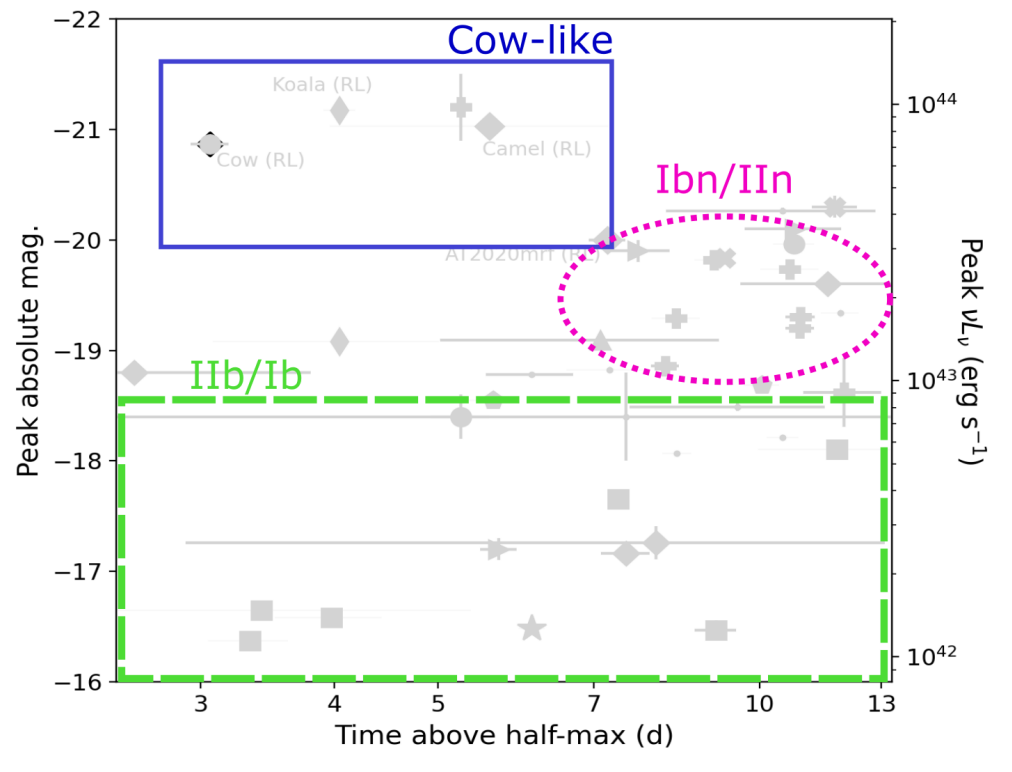}
    \includegraphics[width=0.51\textwidth]{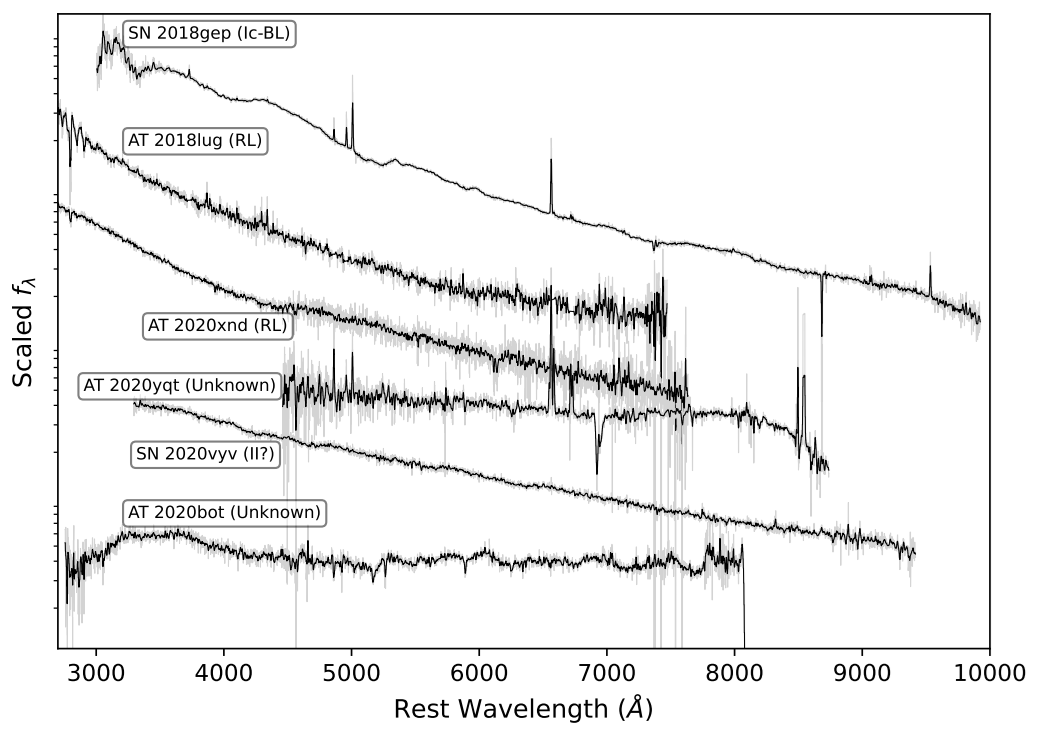}
    \caption{Left: timescale versus peak absolute magnitude parameter space for extragalactic transients, with `Cow-like' LFBOTs indicated in the top left \cite{2023ApJ...949..120H}. The other transients in this parameter space can appear as FBOTs (rather than LFBOTs). Right: a selection of FBOT optical spectra \cite{2023ApJ...949..120H}. }
    \label{fig:fbots}
\end{figure*}

\newpage

\section{Technology and Data Handling Requirements}
\label{sec:tech}
{\bf What is needed:} The depth, cadence and field of view of photometric surveys is ever growing - the newly operational Vera Rubin Observatory is expected to issue millions of alerts per night. While photometric cuts can be made to isolate FBOTs, many transients can appear superficially similar based on these criteria alone, from young/interacting supernovae to LFBOTs to kilonovae to Galactic sources (e.g. cataclysmic variables). With progress in survey {\em detection} likely to continue, a step up is needed in {\em spectroscopic} classification capabilities. So far, the process from target selection (based on photometry) to the triggering of initial spectroscopy to the eventual deployment of large (multi-wavelength) follow-up campaigns is only semi-automated and takes at least a few days, so the crucial early phases are often missed. With orders of magnitude more FBOT candidates being discovered, tools for rapid, automated, same-night photometric classification from transient streams will be essential. Similarly, automated spectroscopic classification from initial follow-up efforts (e.g. with machine learning) will be essential for prompt decisions regarding which objects to (continue) following up. We are currently missing the vast majority of these events: based on the rates presented by \cite{2023ApJ...949..120H}, there could be hundreds per year which peak at optical apparent magnitudes of $\sim$20 or brighter. However, they are are by definition fast evolving, so detection and classification capabilities down to much fainter magnitudes are needed to ensure their detection even if the peak is missed (and to follow their evolution). In particular, the capability for good (S/N $>$10) spectroscopy (and polarimetry, see below) for sources down to $\sim$23$^{\rm rd}$ magnitude would be essential. LFBOTs have featureless spectra at peak, so dedicated follow-up is crucial to capture the emergence of spectral features.

%For FBOTs, rapid spectroscopic classification of a large number of candidates every night is key: these sources fade fast (within days), and rapid classification enables the efficient allocation of resources for further follow-up. Early spectroscopy (ideally same night as discovery) also enables the highest S/N and the ability to classify more distant objects, increasing the volume within which we can probe FBOT demographics, and the evolution of this population over cosmic history.

{\bf Further requirements:}
{\bf Polarimetry} allows us to investigate the geometry of the emitting region throughout the evolution (e.g. \cite{2023MNRAS.521.3323M} found brief phases of high polarisation). Otherwise, and as also found by \cite{2025MNRAS.537.3298P}, LFBOTs seem consistent with high spherical symmetry. This is suggestive of disc-powered systems with spherical outflows. The discovery of minute-long optical flares after the LFBOT AT2022tsd \cite{2023Natur.623..927H}, was possible thanks to {\bf high-cadence, short-exposure photometry}, and raises the prospect of extending the parameter space in Figure 1 down to minute durations or shorter. Might there be undiscovered fast phenomena, perhaps associated with highly variable accretion onto the central engines of extreme transients? \cite{2023Natur.623..927H} note that such variability may only be observable with favourable viewing angles (i.e. clear sight-lines down to the compact object), so high-speed photometric observations could also provide insight into the geometry of LFBOTs, and engine-powered transients more generally.

\printbibliography

@ARTICLE{2023MNRAS.521.3323M,
       author = {{Maund}, Justyn R. and {H{\"o}flich}, Peter A. and {Steele}, Iain A. and {Yang(杨轶)}, Yi and {Wiersema}, Klaas and {Kobayashi}, Shiho and {Jordana-Mitjans}, Nuria and {Mundell}, Carole and {Gomboc}, Andreja and {Guidorzi}, Cristiano and {Smith}, Robert J.},
        title = "{A flash of polarized optical light points to an aspherical 'cow'}",
      journal = {\mnras},
         year = 2023,
        month = may,
       volume = {521},
       number = {3},
        pages = {3323-3332},
          doi = {10.1093/mnras/stad539},
archivePrefix = {arXiv},
       eprint = {2303.00787},
 primaryClass = {astro-ph.SR},
       adsurl = {https://ui.adsabs.harvard.edu/abs/2023MNRAS.521.3323M}
}

@ARTICLE{2023ApJ...949..120H,
       author = {{Ho}, Anna Y.~Q. and {Perley}, Daniel A. and {Gal-Yam}, Avishay and {Lunnan}, Ragnhild and {Sollerman}, Jesper and {Schulze}, Steve and {Das}, Kaustav K. and {Dobie}, Dougal and {Yao}, Yuhan and {Fremling}, Christoffer and {Adams}, Scott and {Anand}, Shreya and {Andreoni}, Igor and {Bellm}, Eric C. and {Bruch}, Rachel J. and {Burdge}, Kevin B. and {Castro-Tirado}, Alberto J. and {Dahiwale}, Aishwarya and {De}, Kishalay and {Dekany}, Richard and {Drake}, Andrew J. and {Duev}, Dmitry A. and {Graham}, Matthew J. and {Helou}, George and {Kaplan}, David L. and {Karambelkar}, Viraj and {Kasliwal}, Mansi M. and {Kool}, Erik C. and {Kulkarni}, S.~R. and {Mahabal}, Ashish A. and {Medford}, Michael S. and {Miller}, A.~A. and {Nordin}, Jakob and {Ofek}, Eran and {Petitpas}, Glen and {Riddle}, Reed and {Sharma}, Yashvi and {Smith}, Roger and {Stewart}, Adam J. and {Taggart}, Kirsty and {Tartaglia}, Leonardo and {Tzanidakis}, Anastasios and {Winters}, Jan Martin},
        title = "{A Search for Extragalactic Fast Blue Optical Transients in ZTF and the Rate of AT2018cow-like Transients}",
      journal = {\apj},
         year = 2023,
        month = jun,
       volume = {949},
       number = {2},
          eid = {120},
        pages = {120},
          doi = {10.3847/1538-4357/acc533},
archivePrefix = {arXiv},
       eprint = {2105.08811},
 primaryClass = {astro-ph.HE},
       adsurl = {https://ui.adsabs.harvard.edu/abs/2023ApJ...949..120H}
}

@ARTICLE{2025ApJ...993...76O,
       author = {{Ofek}, Eran O. and {Ozer}, Lior and {Konno}, Ruslan and {Strasman}, Nimrod and {Chen}, Ping and {Ben-Ami}, Sagi and {Polishook}, David and {Krassilchtchikov}, Alexander and {Garrappa}, Simone and {Zimmermann}, Erez A. and {Segre}, Enrico and {Horowicz}, Asaf and {Gal-Yam}, Avishay and {Shani}, Yarin M. and {Fainer}, Stanislav and {Engel}, Michael and {Sofer-Rimalt}, Yahel and {Ho}, Anna Y.~Q. and {Shvartzvald}, Yossi and {Yaron}, Ofer and {Rybicki}, Kris and {Blumenzweig}, Arie and {Spitzer}, Sarah and {Arad}, Ron},
        title = "{A Search for Minute-timescale Flares from the Transient AT 2024wpp}",
      journal = {\apj},
         year = 2025,
        month = nov,
       volume = {993},
       number = {1},
          eid = {76},
        pages = {76},
          doi = {10.3847/1538-4357/ae00c0},
archivePrefix = {arXiv},
       eprint = {2508.18359},
 primaryClass = {astro-ph.HE},
       adsurl = {https://ui.adsabs.harvard.edu/abs/2025ApJ...993...76O}
}

@ARTICLE{2023Natur.623..927H,
       author = {{Ho}, Anna Y.~Q. and {Perley}, Daniel A. and {Chen}, Ping and {Schulze}, Steve and {Dhillon}, Vik and {Kumar}, Harsh and {Suresh}, Aswin and {Swain}, Vishwajeet and {Bremer}, Michael and {Smartt}, Stephen J. and {Anderson}, Joseph P. and {Anupama}, G.~C. and {Awiphan}, Supachai and {Barway}, Sudhanshu and {Bellm}, Eric C. and {Ben-Ami}, Sagi and {Bhalerao}, Varun and {de Boer}, Thomas and {Brink}, Thomas G. and {Burruss}, Rick and {Chandra}, Poonam and {Chen}, Ting-Wan and {Chen}, Wen-Ping and {Cooke}, Jeff and {Coughlin}, Michael W. and {Das}, Kaustav K. and {Drake}, Andrew J. and {Filippenko}, Alexei V. and {Freeburn}, James and {Fremling}, Christoffer and {Fulton}, Michael D. and {Gal-Yam}, Avishay and {Galbany}, Llu{\'\i}s and {Gao}, Hua and {Graham}, Matthew J. and {Gromadzki}, Mariusz and {Guti{\'e}rrez}, Claudia P. and {Hinds}, K. -Ryan and {Inserra}, Cosimo and {A J}, Nayana and {Karambelkar}, Viraj and {Kasliwal}, Mansi M. and {Kulkarni}, Shri and {M{\"u}ller-Bravo}, Tom{\'a}s E. and {Magnier}, Eugene A. and {Mahabal}, Ashish A. and {Moore}, Thomas and {Ngeow}, Chow-Choong and {Nicholl}, Matt and {Ofek}, Eran O. and {Omand}, Conor M.~B. and {Onori}, Francesca and {Pan}, Yen-Chen and {Pessi}, Priscila J. and {Petitpas}, Glen and {Polishook}, David and {Poshyachinda}, Saran and {Pursiainen}, Miika and {Riddle}, Reed and {Rodriguez}, Antonio C. and {Rusholme}, Ben and {Segre}, Enrico and {Sharma}, Yashvi and {Smith}, Ken W. and {Sollerman}, Jesper and {Srivastav}, Shubham and {Strotjohann}, Nora Linn and {Suhr}, Mark and {Svinkin}, Dmitry and {Wang}, Yanan and {Wiseman}, Philip and {Wold}, Avery and {Yang}, Sheng and {Yang}, Yi and {Yao}, Yuhan and {Young}, David R. and {Zheng}, WeiKang},
        title = "{Minutes-duration optical flares with supernova luminosities}",
      journal = {\nat},
         year = 2023,
        month = nov,
       volume = {623},
       number = {7989},
        pages = {927-931},
          doi = {10.1038/s41586-023-06673-6},
archivePrefix = {arXiv},
       eprint = {2311.10195},
 primaryClass = {astro-ph.HE},
       adsurl = {https://ui.adsabs.harvard.edu/abs/2023Natur.623..927H}
}

@ARTICLE{2025ApJ...993L...6N,
       author = {{Nayana}, A.~J. and {Margutti}, Raffaella and {Wiston}, Eli and {Laskar}, Tanmoy and {Migliori}, Giulia and {Chornock}, Ryan and {Galvin}, Timothy J. and {LeBaron}, Natalie and {Hajela}, Aprajita and {Christy}, Collin T. and {Sfaradi}, Itai and {Tsuna}, Daichi and {Aspegren}, Olivia and {De Colle}, Fabio and {Metzger}, Brian D. and {Lu}, Wenbin and {Beniamini}, Paz and {Kasen}, Daniel and {Berger}, Edo and {Grefenstette}, Brian W. and {Alexander}, Kate D. and {Anupama}, G.~C. and {Coppejans}, Deanne L. and {Cruz}, Luigi F. and {DeBoer}, David R. and {Drout}, Maria R. and {Farah}, Wael and {Huang}, Xiaoshan and {Jacobson-Gal{\'a}n}, W.~V. and {Milisavljevic}, Dan and {Pollak}, Alexander W. and {Roth}, Nathan J. and {Sears}, Huei and {Siemion}, Andrew and {Sheikh}, Sofia Z. and {Steiner}, James F. and {Vurm}, Indrek},
        title = "{The Most Luminous Known Fast Blue Optical Transient AT 2024wpp: Unprecedented Evolution and Properties in the X-Rays and Radio}",
      journal = {\apjl},
         year = 2025,
        month = nov,
       volume = {993},
       number = {1},
          eid = {L6},
        pages = {L6},
          doi = {10.3847/2041-8213/ae0b4d},
archivePrefix = {arXiv},
       eprint = {2509.00952},
 primaryClass = {astro-ph.HE},
       adsurl = {https://ui.adsabs.harvard.edu/abs/2025ApJ...993L...6N}
}

@ARTICLE{2024ApJ...977..162G,
       author = {{Guti{\'e}rrez}, Claudia P. and {Mattila}, Seppo and {Lundqvist}, Peter and {Dessart}, Luc and {Gonz{\'a}lez-Gait{\'a}n}, Santiago and {Jonker}, Peter G. and {Dong}, Subo and {Coppejans}, Deanne and {Chen}, Ping and {Charalampopoulos}, Panos and {Elias-Rosa}, Nancy and {Reynolds}, Thomas M. and {Kochanek}, Christopher and {Fraser}, Morgan and {Pastorello}, Andrea and {Gromadzki}, Mariusz and {Neustadt}, Jack and {Benetti}, Stefano and {Kankare}, Erkki and {Kangas}, Tuomas and {Kotak}, Rubina and {Stritzinger}, Maximilian D. and {Wevers}, Thomas and {Zhang}, Bing and {Bersier}, David and {Bose}, Subhash and {Buckley}, David A.~H. and {Dastidar}, Raya and {Gangopadhyay}, Anjasha and {Hamanowicz}, Aleksandra and {Kollmeier}, Juna A. and {Mao}, Jirong and {Misra}, Kuntal and {Potter}, Stephen. B. and {Prieto}, Jose L. and {Romero-Colmenero}, Encarni and {Singh}, Mridweeka and {Somero}, Auni and {Terreran}, Giacomo and {Vaisanen}, Petri and {Wyrzykowski}, {\L}ukasz},
        title = "{CSS 161010: A Luminous Fast Blue Optical Transient with Broad Blueshifted Hydrogen Lines}",
      journal = {\apj},
         year = 2024,
        month = dec,
       volume = {977},
       number = {2},
          eid = {162},
        pages = {162},
          doi = {10.3847/1538-4357/ad89a5},
archivePrefix = {arXiv},
       eprint = {2408.04698},
 primaryClass = {astro-ph.HE},
       adsurl = {https://ui.adsabs.harvard.edu/abs/2024ApJ...977..162G}
}

@ARTICLE{2025arXiv251003402C,
       author = {{Chrimes}, A.~A. and {Jonker}, P.~G. and {Levan}, A.~J. and {Mummery}, A.},
        title = "{Luminous Fast Blue Optical Transients as very massive star core-collapse events}",
         year = 2025,
        month = oct,
          eid = {arXiv:2510.03402},
        pages = {arXiv:2510.03402},
          doi = {10.48550/arXiv.2510.03402},
archivePrefix = {arXiv},
       eprint = {2510.03402},
 primaryClass = {astro-ph.HE},
       adsurl = {https://ui.adsabs.harvard.edu/abs/2025arXiv251003402C}
}

@ARTICLE{2025arXiv251009745K,
       author = {{Klencki}, Jakub and {Metzger}, Brian D.},
        title = "{Luminous Fast Blue Optical Transients as ``Failed'' Gravitational Wave Sources: Helium Core$-$Black Hole Mergers Following Delayed Dynamical Instability}",
         year = 2025,
        month = oct,
          eid = {arXiv:2510.09745},
        pages = {arXiv:2510.09745},
          doi = {10.48550/arXiv.2510.09745},
archivePrefix = {arXiv},
       eprint = {2510.09745},
 primaryClass = {astro-ph.HE},
       adsurl = {https://ui.adsabs.harvard.edu/abs/2025arXiv251009745K}
}

@ARTICLE{2025MNRAS.537.3298P,
       author = {{Pursiainen}, M. and {Killestein}, T.~L. and {Kuncarayakti}, H. and {Charalampopoulos}, P. and {Warwick}, B. and {Lyman}, J. and {Kotak}, R. and {Leloudas}, G. and {Coppejans}, D. and {Kravtsov}, T. and {Maeda}, K. and {Nagao}, T. and {Taguchi}, K. and {Ackley}, K. and {Dhillon}, V.~S. and {Galloway}, D.~K. and {Kumar}, A. and {O'Neill}, D. and {Ramsay}, G. and {Steeghs}, D.},
        title = "{Optical evolution of AT 2024wpp: the high-velocity outflows in Cow-like transients are consistent with high spherical symmetry}",
      journal = {\mnras},
         year = 2025,
        month = mar,
       volume = {537},
       number = {4},
        pages = {3298-3309},
          doi = {10.1093/mnras/staf232},
archivePrefix = {arXiv},
       eprint = {2411.03272},
 primaryClass = {astro-ph.HE},
       adsurl = {https://ui.adsabs.harvard.edu/abs/2025MNRAS.537.3298P}
}

@ARTICLE{2019ApJ...872...18M,
       author = {{Margutti}, R. and {Metzger}, B.~D. and {Chornock}, R. and {Vurm}, I. and {Roth}, N. and {Grefenstette}, B.~W. and {Savchenko}, V. and {Cartier}, R. and {Steiner}, J.~F. and {Terreran}, G. and {Margalit}, B. and {Migliori}, G. and {Milisavljevic}, D. and {Alexander}, K.~D. and {Bietenholz}, M. and {Blanchard}, P.~K. and {Bozzo}, E. and {Brethauer}, D. and {Chilingarian}, I.~V. and {Coppejans}, D.~L. and {Ducci}, L. and {Ferrigno}, C. and {Fong}, W. and {G{\"o}tz}, D. and {Guidorzi}, C. and {Hajela}, A. and {Hurley}, K. and {Kuulkers}, E. and {Laurent}, P. and {Mereghetti}, S. and {Nicholl}, M. and {Patnaude}, D. and {Ubertini}, P. and {Banovetz}, J. and {Bartel}, N. and {Berger}, E. and {Coughlin}, E.~R. and {Eftekhari}, T. and {Frederiks}, D.~D. and {Kozlova}, A.~V. and {Laskar}, T. and {Svinkin}, D.~S. and {Drout}, M.~R. and {MacFadyen}, A. and {Paterson}, K.},
        title = "{An Embedded X-Ray Source Shines through the Aspherical AT 2018cow: Revealing the Inner Workings of the Most Luminous Fast-evolving Optical Transients}",
      journal = {\apj},
         year = 2019,
        month = feb,
       volume = {872},
       number = {1},
          eid = {18},
        pages = {18},
          doi = {10.3847/1538-4357/aafa01},
archivePrefix = {arXiv},
       eprint = {1810.10720},
 primaryClass = {astro-ph.HE},
       adsurl = {https://ui.adsabs.harvard.edu/abs/2019ApJ...872...18M}
}

@ARTICLE{2022NatAs...6..249P,
       author = {{Pasham}, Dheeraj R. and {Ho}, Wynn C.~G. and {Alston}, William and {Remillard}, Ronald and {Ng}, Mason and {Gendreau}, Keith and {Metzger}, Brian D. and {Altamirano}, Diego and {Chakrabarty}, Deepto and {Fabian}, Andrew and {Miller}, Jon and {Bult}, Peter and {Arzoumanian}, Zaven and {Steiner}, James F. and {Strohmayer}, Tod and {Tombesi}, Francesco and {Homan}, Jeroen and {Cackett}, Edward M. and {Harding}, Alice},
        title = "{Evidence for a compact object in the aftermath of the extragalactic transient AT2018cow}",
      journal = {Nature Astronomy},
         year = 2021,
        month = dec,
       volume = {6},
        pages = {249-258},
          doi = {10.1038/s41550-021-01524-8},
archivePrefix = {arXiv},
       eprint = {2112.04531},
 primaryClass = {astro-ph.HE},
       adsurl = {https://ui.adsabs.harvard.edu/abs/2022NatAs...6..249P}
}

@ARTICLE{2019MNRAS.484.1031P,
       author = {{Perley}, Daniel A. and {Mazzali}, Paolo A. and {Yan}, Lin and {Cenko}, S. Bradley and {Gezari}, Suvi and {Taggart}, Kirsty and {Blagorodnova}, Nadia and {Fremling}, Christoffer and {Mockler}, Brenna and {Singh}, Avinash and {Tominaga}, Nozomu and {Tanaka}, Masaomi and {Watson}, Alan M. and {Ahumada}, Tom{\'a}s and {Anupama}, G.~C. and {Ashall}, Chris and {Becerra}, Rosa L. and {Bersier}, David and {Bhalerao}, Varun and {Bloom}, Joshua S. and {Butler}, Nathaniel R. and {Copperwheat}, Chris and {Coughlin}, Michael W. and {De}, Kishalay and {Drake}, Andrew J. and {Duev}, Dmitry A. and {Frederick}, Sara and {Gonz{\'a}lez}, J. Jes{\'u}s and {Goobar}, Ariel and {Heida}, Marianne and {Ho}, Anna Y.~Q. and {Horst}, John and {Hung}, Tiara and {Itoh}, Ryosuke and {Jencson}, Jacob E. and {Kasliwal}, Mansi M. and {Kawai}, Nobuyuki and {Khanam}, Tanazza and {Kulkarni}, Shrinivas R. and {Kumar}, Brajesh and {Kumar}, Harsh and {Kutyrev}, Alexander S. and {Lee}, William H. and {Maeda}, Keiichi and {Mahabal}, Ashish and {Murata}, Katsuhiro L. and {Neill}, James D. and {Ngeow}, Chow-Choong and {Penprase}, Bryan and {Pian}, Elena and {Quimby}, Robert and {Ramirez-Ruiz}, Enrico and {Richer}, Michael G. and {Rom{\'a}n-Z{\'u}{\~n}iga}, Carlos G. and {Sahu}, D.~K. and {Srivastav}, Shubham and {Socia}, Quentin and {Sollerman}, Jesper and {Tachibana}, Yutaro and {Taddia}, Francesco and {Tinyanont}, Samaporn and {Troja}, Eleonora and {Ward}, Charlotte and {Wee}, Jerrick and {Yu}, Po-Chieh},
        title = "{The fast, luminous ultraviolet transient AT2018cow: extreme supernova, or disruption of a star by an intermediate-mass black hole?}",
      journal = {\mnras},
         year = 2019,
        month = mar,
       volume = {484},
       number = {1},
        pages = {1031-1049},
          doi = {10.1093/mnras/sty3420},
archivePrefix = {arXiv},
       eprint = {1808.00969},
 primaryClass = {astro-ph.HE},
       adsurl = {https://ui.adsabs.harvard.edu/abs/2019MNRAS.484.1031P}
}

@ARTICLE{2018PASP..130f4505T,
       author = {{Tonry}, J.~L. and {Denneau}, L. and {Heinze}, A.~N. and {Stalder}, B. and {Smith}, K.~W. and {Smartt}, S.~J. and {Stubbs}, C.~W. and {Weiland}, H.~J. and {Rest}, A.},
        title = "{ATLAS: A High-cadence All-sky Survey System}",
      journal = {\pasp},
         year = 2018,
        month = jun,
       volume = {130},
       number = {988},
        pages = {064505},
          doi = {10.1088/1538-3873/aabadf},
archivePrefix = {arXiv},
       eprint = {1802.00879},
 primaryClass = {astro-ph.IM},
       adsurl = {https://ui.adsabs.harvard.edu/abs/2018PASP..130f4505T}
}

%\begin{multicols}{2}
%\bibliographystyle{plain}
%{\footnotesize
%\bibliography{bib} % create refs.bib if needed
%\end{multicols}

\end{document}